\documentclass[useAMS,usenatbib]{mn2e}

\usepackage{graphicx}
\title[Electron impact excitation of Be-like ions]{Electron impact excitation of Be-like ions: a comparison of DARC and ICFT results}
\author[K. M. Aggarwal and F. P. Keenan]{Kanti  M.  ~Aggarwal$^{1}$\thanks{E-mail:
 K.Aggarwal@qub.ac.uk(KMA); F.Keenan@qub.ac.uk (FPK)} and Francis  P.   ~Keenan$^{1}$ \\
$^{1}$Astrophysics Research Centre, School of Mathematics and Physics, Queen's University Belfast, Belfast BT7 1NN,  UK} %\\
\begin{document}

\date{Accepted 2015 December 18. Received 2014 December 11; in original form 2014 November 5 }

\pagerange{\pageref{firstpage}--\pageref{lastpage}} \pubyear{2015}

\maketitle

\label{firstpage}

\begin{abstract}
Emission lines of Be-like ions are frequently observed in astrophysical plasmas, and many  are useful for density and temperature diagnostics. However, accurate atomic data for  energy levels, radiative rates (A-values) and  effective electron excitation collision strengths ($\Upsilon$) are required for reliable plasma modelling. In general it is reasonably straightforward to calculate  energy levels and A- values  to a high level of accuracy. By contrast, considerable effort is required to calculate $\Upsilon$, and hence it is not always possible to assess the accuracy of available data. Recently,  two independent calculations (adopting  the $R$-matrix method) but with different approaches (DARC and ICFT) have appeared for a range of Be-like ions. Therefore, in this work we compare the two sets of $\Upsilon$, highlight the large discrepancies for a significant number of transitions  and suggest possible reasons for these. 

\end{abstract}


\section{Introduction}
Emission lines of Be-like ions are widely detected in a variety of astrophysical plasmas, including the solar transition region and corona, with the prominent ions including  C III, O V, Ca XVII and Fe XXIII -- see for example, the CHIANTI database at {\tt http://www.chiantidatabase.org/} and  the {\em Atomic Line List} (v2.04) of Peter van Hoof at ${\tt {\verb+http://www.pa.uky.edu/~peter/atomic/+}}$. Many of the observed lines are sensitive to variations in  density or temperature, and hence are useful as diagnostics -- see for example, \cite{el1}.  However, for the reliable modelling of plasmas, accurate atomic data are required, particularly for energy levels, radiative rates (A-values), and excitation rates or equivalently the effective collision strengths ($\Upsilon$), which are obtained from the electron impact collision strengths ($\Omega$). Unfortunately, existing atomic data for Be-like ions prior to 2014 were very limited, particularly for $\Upsilon$. However, very recently  \cite*{icft} have reported data for a range of Be-like ions up to Z = 36. For their calculations they have adopted the {\em AutoStructure} (AS) code of \cite{as} for the generation of wave functions, i.e. to determine energy levels and A-values. In the subsequent calculations of $\Omega$ and $\Upsilon$ they have adopted the $R$-matrix code of \cite*{rm2}. Their $\Omega$ are primarily obtained in  the $LS$ coupling (Russell-Saunders or spin-orbit coupling) and corresponding results for {\em fine-structure} transitions  are determined  through their intermediate coupling frame transformation (ICFT) method.

Assessing atomic data, particularly for $\Upsilon$, is a difficult task \citep{fst} mainly because the calculations are highly complicated, require large computational resources and cannot be easily repeated. Therefore, the assessment is frequently based on limited comparisons either with already available results or with simple calculations, such as with the {\em distorted-wave} method but without resonances. However, often such accuracy assessments are far from satisfactory as noted by \cite{fst} for several ions. 

\begin{table*}
 \centering
% \begin{minipage}{140mm}
  \caption{Energies for the lowest 80 levels of Al X (in Ryd).}
 \begin{tabular}{rllrrrllrrrr} \hline
Index  & Configuration  & Level   & GRASP        &  AS   & Index  & Configuration  & Level   & GRASP        &  AS   \\
 \hline
   1 &  2s$^2$    &  $^1$S$  _0$  &   0.00000	 &   0.00000   &     41 &  2p3d      &  $^3$P$^o_2$  &  19.77978    &   19.77212     \\
   2 &  2s2p	  &  $^3$P$^o_0$  &   1.41897	 &   1.41738   &     42 &  2p3d      &  $^3$P$^o_1$  &  19.79099    &   19.78275     \\ 
   3 &  2s2p	  &  $^3$P$^o_1$  &   1.43387	 &   1.43437   &     43 &  2p3d      &  $^3$P$^o_0$  &  19.79743    &   19.78828     \\ 
   4 &  2s2p	  &  $^3$P$^o_2$  &   1.46662	 &   1.46968   &     44 &  2p3p      &  $^1$S$  _0$  &  19.94280    &   19.92116     \\ 
   5 &  2s2p	  &  $^1$P$^o_1$  &   2.81633	 &   2.79949   &     45 &  2p3d      &  $^1$F$^o_3$  &  20.02755    &   20.01046     \\ 
   6 &  2p$^2$    &  $^3$P$  _0$  &   3.71907	 &   3.71638   &     46 &  2p3d      &  $^1$P$^o_1$  &  20.07597    &   20.06076     \\ 
   7 &  2p$^2$    &  $^3$P$  _1$  &   3.73652	 &   3.73517   &     47 &  2s4s      &  $^3$S$  _1$  &  22.54097    &   22.53181     \\ 
   8 &  2p$^2$    &  $^3$P$  _2$  &   3.76518	 &   3.76832   &     48 &  2s4s      &  $^1$S$  _0$  &  22.64271    &   22.63203     \\
   9 &  2p$^2$    &  $^1$D$  _2$  &   4.17504	 &   4.17207   &     49 &  2s4p      &  $^3$P$^o_0$  &  22.79178    &   22.78517     \\
  10 &  2p$^2$    &  $^1$S$  _0$  &   5.17938	 &   5.16932   &     50 &  2s4p      &  $^3$P$^o_1$  &  22.79331    &   22.78674     \\
  11 &  2s3s	  &  $^3$S$  _1$  &  16.89542	 &  16.88436   &     51 &  2s4p      &  $^3$P$^o_2$  &  22.79723    &   22.79057     \\
  12 &  2s3s	  &  $^1$S$  _0$  &  17.15693	 &  17.15202   &     52 &  2s4p      &  $^1$P$^o_1$  &  22.82350    &   22.81370     \\
  13 &  2s3p	  &  $^1$P$^o_1$  &  17.52796	 &  17.51781   &     53 &  2s4d      &  $^3$D$  _1$  &  22.93591    &   22.92974     \\
  14 &  2s3p	  &  $^3$P$^o_0$  &  17.55008	 &  17.54198   &     54 &  2s4d      &  $^3$D$  _2$  &  22.93645    &   22.93041     \\
  15 &  2s3p	  &  $^3$P$^o_1$  &  17.55555	 &  17.54746   &     55 &  2s4d      &  $^3$D$  _3$  &  22.93741    &   22.93144     \\
  16 &  2s3p	  &  $^3$P$^o_2$  &  17.56323	 &  17.55522   &     56 &  2s4d      &  $^1$D$  _2$  &  23.02777    &   23.01752     \\
  17 &  2s3d	  &  $^3$D$  _1$  &  17.90599	 &  17.89877   &     57 &  2s4f      &  $^3$F$^o_2$  &  23.02937    &   23.02096     \\ 
  18 &  2s3d	  &  $^3$D$  _2$  &  17.90756	 &  17.90073   &     58 &  2s4f      &  $^3$F$^o_3$  &  23.02961    &   23.02133     \\
  19 &  2s3d	  &  $^3$D$  _3$  &  17.91022	 &  17.90367   &     59 &  2s4f      &  $^3$F$^o_4$  &  23.03003    &   23.02184     \\
  20 &  2s3d	  &  $^1$D$  _2$  &  18.17882	 &  18.16519   &     60 &  2s4f      &  $^1$F$^o_3$  &  23.05392    &   23.04587     \\
  21 &  2p3s	  &  $^3$P$^o_0$  &  18.69110	 &  18.68325   &     61 &  2p4s      &  $^3$P$^o_0$  &  24.20663    &   24.20036     \\
  22 &  2p3s	  &  $^3$P$^o_1$  &  18.70615	 &  18.69882   &     62 &  2p4s      &  $^3$P$^o_1$  &  24.21768    &   24.21073     \\
  23 &  2p3s	  &  $^3$P$^o_2$  &  18.74171	 &  18.73549   &     63 &  2p4s      &  $^3$P$^o_2$  &  24.25816    &   24.25301     \\
  24 &  2p3s	  &  $^1$P$^o_1$  &  19.00401	 &  18.99133   &     64 &  2p4s      &  $^1$P$^o_1$  &  24.32131    &   24.30294     \\
  25 &  2p3p	  &  $^1$P$  _1$  &  19.08552	 &  19.07491   &     65 &  2p4p      &  $^1$P$  _1$  &  24.38599    &   24.37956     \\
  26 &  2p3p	  &  $^3$D$  _1$  &  19.16008	 &  19.14799   &     66 &  2p4p      &  $^3$D$  _1$  &  24.41706    &   24.41031     \\
  27 &  2p3p	  &  $^3$D$  _2$  &  19.17515	 &  19.16319   &     67 &  2p4p      &  $^3$D$  _2$  &  24.42038    &   24.41402     \\
  28 &  2p3p	  &  $^3$D$  _3$  &  19.20844	 &  19.19688   &     68 &  2p4p      &  $^3$D$  _3$  &  24.45399    &   24.44816     \\
  29 &  2p3p	  &  $^3$S$  _1$  &  19.32141	 &  19.30707   &     69 &  2p4p      &  $^3$S$  _1$  &  24.47906    &   24.46575     \\
  30 &  2p3p	  &  $^3$P$  _0$  &  19.38473	 &  19.37790   &     70 &  2p4p      &  $^3$P$  _0$  &  24.48248    &   24.47470     \\
  31 &  2p3p	  &  $^3$P$  _1$  &  19.40025	 &  19.39402   &     71 &  2p4p      &  $^3$P$  _1$  &  24.50993    &   24.50041     \\
  32 &  2p3p	  &  $^3$P$  _2$  &  19.41709	 &  19.41265   &     72 &  2p4p      &  $^3$P$  _2$  &  24.51423    &   24.50850     \\
  33 &  2p3d	  &  $^3$F$^o_2$  &  19.46968	 &  19.46172   &     73 &  2p4d      &  $^3$F$^o_2$  &  24.53225    &   24.52677     \\
  34 &  2p3d	  &  $^3$F$^o_3$  &  19.49594	 &  19.48869   &     74 &  2p4d      &  $^3$F$^o_3$  &  24.55592    &   24.55041     \\
  35 &  2p3d	  &  $^1$D$^o_2$  &  19.50998	 &  19.50530   &     75 &  2p4d      &  $^1$D$^o_2$  &  24.56504    &   24.56022     \\
  36 &  2p3d	  &  $^3$F$^o_4$  &  19.52225	 &  19.51643   &     76 &  2p4p      &  $^1$D$  _2$  &  24.58233    &   24.57285     \\
  37 &  2p3p	  &  $^1$D$  _2$  &  19.60698	 &  19.59904   &     77 &  2p4d      &  $^3$F$^o_4$  &  24.58565    &   24.58108     \\
  38 &  2p3d	  &  $^3$D$^o_1$  &  19.68861	 &  19.68287   &     78 &  2p4d      &  $^3$D$^o_1$  &  24.61082    &   24.60428     \\
  39 &  2p3d	  &  $^3$D$^o_2$  &  19.69591	 &  19.69032   &     79 &  2p4d      &  $^3$D$^o_2$  &  24.62033    &   24.61401     \\
  40 &  2p3d	  &  $^3$D$^o_3$  &  19.70995	 &  19.70524   &     80 &  2p4f      &  $^1$F$  _3$  &  24.62916    &   24.62052     \\
\hline	
\end{tabular}

%\vspace*{0.5 cm}
\begin{flushleft}
{\small
%NIST: {\tt http://www.nist.gov/pml/data/asd.cfm} \\
GRASP: \cite{alx} \\ %Aggarwal \& Keenan (2014)\\
AS: \cite{icft} \\ %Fern{\'a}ndez-Menchero et al. (2014) \\
}
\end{flushleft}
\end{table*}

Realising the importance of Be-like ions we have also reported atomic data for several, namely Al X \citep{alx}, Cl XIV, K XVI and Ge XXIX \citep{cl14} and Ti XIX \citep{ti19}. Our calculations are  independent of those of  \cite{icft} and are based on the {\em Dirac Atomic $R$-matrix Code} ({\sc darc}) of P. H. Norrington and I. P. Grant ({\tt http://web.am.qub.ac.uk/DARC/}). Unlike the semi-relativistic version of the standard $R$-matrix code \citep{rm2} adopted by \cite{icft}, DARC  is based on the $jj$ coupling scheme. The accuracy of the data calculated  (for $\Omega$ and subsequently $\Upsilon$) through this approach is generally higher, because resonances through the energies of degenerating levels are also taken into  account.  For this reason it particularly affects  transitions among the {\em fine-structure} levels of a state. The degeneracy among such levels increases with increasing Z -- see for example levels of Ge XXIX in table 3 of \cite{cl14}. 

Since two independent calculations for 5 Be-like ions with 13 $\le$ Z $\le$ 32 using the same $R$-matrix method (although in different approximations) and of similar complexity are now available, it is  possible to make a detailed comparison to assess their accuracy. This is important, given the  large discrepancies recently noted for transitions of Fe XIV between our calculations with DARC \citep{fe14} and those of \cite{gyl} with ICFT. 

\begin{table*}
 \centering
% \begin{minipage}{140mm}
  \caption{Comparison of effective collision strengths ($\Upsilon$) for resonance transitions of Al X. $a{\pm}b \equiv a{\times}$10$^{{\pm}b}$.}
\begin{tabular}{rrlllllllll} \hline
\multicolumn{2}{c}{Transition} & \multicolumn{3}{c}{DARC (log T$_e$, K)} & \multicolumn{3}{c}{ICFT (log T$_e$, K)} \\
\hline
   I  &      J    &      4.30      &     6.00       &     7.30    &   4.30        &     6.00      &    7.30  \\
 \hline
    1  &      2  &    1.751$-$2  &    1.125$-$2  &    2.333$-$3  &    2.06$-$2  &    1.09$-$2  &    2.33$-$3  \\
    1  &      3  &    5.841$-$2  &    3.421$-$2  &    7.510$-$3  &    7.47$-$2  &    3.31$-$2  &    8.33$-$3  \\
    1  &      4  &    1.127$-$1  &    5.695$-$2  &    1.173$-$2  &    1.28$-$1  &    5.43$-$2  &    1.16$-$2  \\
    1  &      5  &    1.128$-$0  &    1.311$-$0  &    1.940$-$0  &    1.13$-$0  &    1.30$-$0  &    2.16$-$0  \\
    1  &      6  &    3.925$-$4  &    6.619$-$4  &    1.485$-$4  &    3.07$-$4  &    5.40$-$4  &    1.20$-$4  \\
    1  &      7  &    1.021$-$3  &    1.871$-$3  &    4.029$-$4  &    1.08$-$3  &    1.59$-$3  &    3.33$-$4  \\
    1  &      8  &    1.975$-$3  &    3.002$-$3  &    6.714$-$4  &    1.95$-$3  &    2.69$-$3  &    6.12$-$4  \\
    1  &      9  &    1.448$-$2  &    1.734$-$2  &    1.286$-$2  &    1.43$-$2  &    1.69$-$2  &    1.34$-$2  \\
    1  &     10  &    5.105$-$3  &    7.309$-$3  &    3.964$-$3  &    4.94$-$3  &    7.47$-$3  &    3.99$-$3  \\
    1  &     11  &    6.103$-$2  &    1.080$-$2  &    1.291$-$3  &    3.97$-$2  &    9.33$-$3  &    1.22$-$3  \\
    1  &     12  &    9.534$-$2  &    6.066$-$2  &    6.852$-$2  &    7.54$-$2  &    5.98$-$2  &    7.14$-$2  \\
    1  &     13  &    4.994$-$2  &    3.739$-$2  &    1.327$-$1  &    3.93$-$2  &    3.66$-$2  &    1.50$-$1  \\
    1  &     14  &    5.753$-$3  &    1.749$-$3  &    2.783$-$4  &    3.89$-$3  &    1.58$-$3  &    2.72$-$4  \\
    1  &     15  &    1.965$-$2  &    7.004$-$3  &    8.022$-$3  &    1.31$-$2  &    5.89$-$3  &    6.37$-$3  \\
    1  &     16  &    3.012$-$2  &    8.781$-$3  &    1.393$-$3  &    2.07$-$2  &    7.92$-$3  &    1.36$-$3  \\
    1  &     17  &    1.232$-$2  &    6.750$-$3  &    1.304$-$3  &    7.94$-$3  &    6.13$-$3  &    1.26$-$3  \\
    1  &     18  &    2.039$-$2  &    1.125$-$2  &    2.179$-$3  &    1.38$-$2  &    1.02$-$2  &    2.23$-$3  \\
    1  &     19  &    2.843$-$2  &    1.578$-$2  &    3.047$-$3  &    1.89$-$2  &    1.43$-$2  &    2.94$-$3  \\
    1  &     20  &    5.683$-$2  &    7.410$-$2  &    1.424$-$1  &    5.58$-$2  &    7.11$-$2  &    1.48$-$1  \\
    1  &     21  &    8.029$-$4  &    1.803$-$4  &    1.379$-$5  &    6.31$-$4  &    1.53$-$4  &    1.26$-$5  \\
    1  &     22  &    2.980$-$3  &    6.267$-$4  &    1.037$-$4  &    2.17$-$3  &    4.80$-$4  &    9.90$-$5  \\
    1  &     23  &    4.853$-$3  &    1.014$-$3  &    7.485$-$5  &    3.91$-$3  &    7.54$-$4  &    6.18$-$5  \\
    1  &     24  &    1.127$-$2  &    2.773$-$3  &    4.427$-$3  &    8.40$-$3  &    2.56$-$3  &    4.60$-$3  \\
    1  &     25  &    3.740$-$3  &    5.209$-$4  &    8.951$-$5  &    4.84$-$3  &    4.82$-$4  &    8.37$-$5  \\
    1  &     26  &    2.525$-$3  &    4.643$-$4  &    6.364$-$5  &    1.20$-$3  &    3.85$-$4  &    5.97$-$5  \\
    1  &     27  &    4.160$-$3  &    7.776$-$4  &    1.041$-$4  &    2.17$-$3  &    6.40$-$4  &    1.03$-$4  \\
    1  &     28  &    4.976$-$3  &    1.026$-$3  &    1.347$-$4  &    3.27$-$3  &    8.73$-$4  &    1.29$-$4  \\
    1  &     29  &    5.235$-$3  &    5.880$-$4  &    5.978$-$5  &    5.09$-$3  &    6.35$-$4  &    6.61$-$5  \\
    1  &     30  &    7.920$-$4  &    1.069$-$4  &    1.160$-$5  &    5.75$-$4  &    6.49$-$5  &    9.34$-$6  \\
    1  &     31  &    4.201$-$3  &    3.219$-$4  &    2.905$-$5  &    2.07$-$3  &    1.96$-$4  &    2.16$-$5  \\
    1  &     32  &    6.179$-$3  &    4.767$-$4  &    5.373$-$5  &    3.55$-$3  &    3.12$-$4  &    4.59$-$5  \\
    1  &     33  &    1.855$-$3  &    5.452$-$4  &    8.664$-$5  &    8.55$-$4  &    4.88$-$4  &    7.97$-$5  \\
    1  &     34  &    1.673$-$3  &    7.038$-$4  &    9.523$-$5  &    1.23$-$3  &    6.45$-$4  &    9.81$-$5  \\
    1  &     35  &    1.645$-$3  &    5.751$-$4  &    1.294$-$4  &    1.72$-$3  &    5.51$-$4  &    1.11$-$4  \\
    1  &     36  &    1.892$-$3  &    8.610$-$4  &    1.157$-$4  &    1.66$-$3  &    8.18$-$4  &    1.16$-$4  \\
    1  &     37  &    1.967$-$3  &    1.238$-$3  &    1.164$-$3  &    1.95$-$3  &    1.23$-$3  &    1.24$-$3  \\
    1  &     38  &    1.888$-$4  &    2.327$-$4  &    8.316$-$5  &    8.24$-$5  &    1.48$-$4  &    8.07$-$5  \\
    1  &     39  &    3.312$-$4  &    3.172$-$4  &    3.247$-$5  &    1.50$-$4  &    2.21$-$4  &    2.61$-$5  \\
    1  &     40  &    4.338$-$4  &    3.794$-$4  &    3.830$-$5  &    1.83$-$4  &    2.57$-$4  &    3.14$-$5  \\
\hline
\end{tabular}   
\end{table*}

%\newpage  
\section[]{Details of Calculation}
In our work  the fully relativistic {\sc grasp} (General-purpose Relativistic Atomic Structure  Package) code is employed to determine the wave functions. There are several versions of this code, but all are based on the one  originally  developed by  \cite{grasp0},  often  referred to as GRASP0.  The version used by us  has been extensively revised by one of its authors (Dr. P. H. Norrington),  is freely  available at the website {\tt http://web.am.qub.ac.uk/DARC/} and yields comparable results for energy levels and A-values as obtained with using other revisions.  It is  fully relativistic,  based on the $jj$ coupling scheme, and includes higher-order relativistic corrections arising from the Breit (magnetic) interaction and quantum electrodynamics effects (vacuum polarisation and Lamb shift). Additionally,  the option of {\em extended average level} has been adopted for all ions, under which  a weighted (proportional to 2$j$+1) trace of the Hamiltonian matrix is minimised.  

\begin{table*}
 \centering
\begin{tabular}{rrlllllllll} \hline
\multicolumn{2}{c}{Transition} & \multicolumn{3}{c}{DARC (log T$_e$, K)} & \multicolumn{3}{c}{ICFT (log T$_e$, K)} \\
\hline
   I  &      J    &      4.30      &     6.00       &     7.30    &   4.30        &     6.00      &    7.30  \\
 \hline    
    1  &     41  &    1.054$-$3  &    7.252$-$4  &    1.069$-$4  &    6.31$-$4  &    7.78$-$4  &    1.13$-$4  \\
    1  &     42  &    6.700$-$4  &    4.711$-$4  &    8.345$-$5  &    3.72$-$4  &    4.84$-$4  &    9.75$-$5  \\
    1  &     43  &    2.164$-$4  &    1.491$-$4  &    2.203$-$5  &    1.23$-$4  &    1.61$-$4  &    2.33$-$5  \\
    1  &     44  &    1.531$-$3  &    9.485$-$4  &    5.759$-$4  &    1.25$-$3  &    1.11$-$3  &    6.90$-$4  \\
    1  &     45  &    9.098$-$4  &    1.390$-$3  &    1.151$-$3  &    8.39$-$4  &    1.44$-$3  &    1.32$-$3  \\
    1  &     46  &    3.728$-$3  &    4.801$-$3  &    9.630$-$3  &    3.65$-$3  &    4.87$-$3  &    1.03$-$2  \\
    1  &     47  &    9.284$-$3  &    1.669$-$3  &    3.098$-$4  &    1.11$-$2  &    1.94$-$3  &    3.26$-$4  \\
    1  &     48  &    1.830$-$2  &    1.185$-$2  &    1.384$-$2  &    2.19$-$2  &    1.15$-$2  &    1.42$-$2  \\
    1  &     49  &    1.342$-$3  &    4.996$-$4  &    9.613$-$5  &    1.87$-$3  &    5.41$-$4  &    9.80$-$5  \\
    1  &     50  &    4.530$-$3  &    1.576$-$3  &    5.461$-$4  &    5.73$-$3  &    1.70$-$3  &    6.36$-$4  \\
    1  &     51  &    6.416$-$3  &    2.463$-$3  &    4.768$-$4  &    9.48$-$3  &    2.71$-$3  &    4.90$-$4  \\
    1  &     52  &    1.088$-$2  &    1.008$-$2  &    3.006$-$2  &    1.53$-$2  &    9.68$-$3  &    3.22$-$2  \\
    1  &     53  &    3.009$-$3  &    1.877$-$3  &    4.081$-$4  &    2.78$-$3  &    1.84$-$3  &    4.04$-$4  \\
    1  &     54  &    4.942$-$3  &    3.125$-$3  &    6.814$-$4  &    4.64$-$3  &    3.07$-$3  &    7.25$-$4  \\
    1  &     55  &    6.894$-$3  &    4.374$-$3  &    9.513$-$4  &    6.46$-$3  &    4.30$-$3  &    9.43$-$4  \\
    1  &     56  &    1.425$-$2  &    1.493$-$2  &    2.732$-$2  &    1.01$-$2  &    1.30$-$2  &    2.67$-$2  \\
    1  &     57  &    1.685$-$3  &    1.100$-$3  &    1.973$-$4  &    1.94$-$3  &    1.27$-$3  &    2.10$-$4  \\
    1  &     58  &    2.613$-$3  &    1.555$-$3  &    2.778$-$4  &    2.71$-$3  &    1.78$-$3  &    3.15$-$4  \\
    1  &     59  &    3.032$-$3  &    1.978$-$3  &    3.547$-$4  &    3.49$-$3  &    2.29$-$3  &    3.78$-$4  \\
    1  &     60  &    6.258$-$3  &    5.002$-$3  &    7.684$-$3  &    4.98$-$3  &    5.34$-$3  &    8.13$-$3  \\
    1  &     61  &    7.022$-$5  &    1.104$-$5  &    1.956$-$6  &    1.84$-$4  &    2.84$-$5  &    3.01$-$6  \\
    1  &     62  &    2.462$-$4  &    4.459$-$5  &    1.894$-$5  &    6.90$-$4  &    1.03$-$4  &    1.61$-$5  \\
    1  &     63  &    3.067$-$4  &    5.292$-$5  &    9.511$-$6  &    1.35$-$3  &    1.42$-$4  &    1.48$-$5  \\
    1  &     64  &    5.487$-$4  &    1.784$-$4  &    1.698$-$4  &    3.12$-$3  &    2.94$-$4  &    9.40$-$5  \\
    1  &     65  &    1.626$-$4  &    4.927$-$5  &    1.409$-$5  &    6.88$-$4  &    1.16$-$4  &    1.77$-$5  \\
    1  &     66  &    1.310$-$4  &    4.662$-$5  &    1.255$-$5  &    7.56$-$4  &    1.20$-$4  &    1.65$-$5  \\
    1  &     67  &    2.124$-$4  &    7.299$-$5  &    1.818$-$5  &    1.37$-$3  &    2.26$-$4  &    2.65$-$5  \\
    1  &     68  &    2.643$-$4  &    9.706$-$5  &    2.190$-$5  &    1.91$-$3  &    3.08$-$4  &    3.25$-$5  \\
    1  &     69  &    1.368$-$4  &    5.386$-$5  &    1.198$-$5  &    7.17$-$4  &    1.43$-$4  &    2.23$-$5  \\
    1  &     70  &    2.675$-$5  &    1.261$-$5  &    5.294$-$6  &    6.99$-$5  &    3.10$-$5  &    2.18$-$5  \\
    1  &     71  &    1.219$-$4  &    4.371$-$5  &    9.004$-$6  &    4.15$-$4  &    7.54$-$5  &    1.11$-$5  \\
    1  &     72  &    1.125$-$4  &    4.630$-$5  &    1.164$-$5  &    4.81$-$4  &    7.88$-$5  &    1.32$-$5  \\
    1  &     73  &    1.545$-$4  &    8.364$-$5  &    2.041$-$5  &    8.13$-$4  &    1.52$-$4  &    2.52$-$5  \\
    1  &     74  &    1.877$-$4  &    1.100$-$4  &    2.396$-$5  &    1.24$-$3  &    2.06$-$4  &    3.66$-$5  \\
    1  &     75  &    1.657$-$4  &    8.476$-$5  &    2.622$-$5  &    1.15$-$3  &    1.51$-$4  &    2.81$-$5  \\
    1  &     76  &    2.350$-$4  &    1.284$-$4  &    1.143$-$4  &    1.59$-$3  &    2.43$-$4  &    9.16$-$5  \\
    1  &     77  &    2.456$-$4  &    1.402$-$4  &    2.784$-$5  &    1.43$-$3  &    2.56$-$4  &    3.85$-$5  \\
    1  &     78  &    7.624$-$5  &    6.637$-$5  &    8.803$-$5  &    3.57$-$4  &    9.66$-$5  &    1.29$-$4  \\
    1  &     79  &    1.111$-$4  &    6.312$-$5  &    1.416$-$5  &    7.50$-$4  &    1.11$-$4  &    1.78$-$5  \\
    1  &     80  &    4.524$-$5  &    2.506$-$5  &    8.451$-$6  &    7.04$-$4  &    8.66$-$5  &    1.09$-$5  \\
\hline
\end{tabular} 

%\vspace*{0.5 cm}
\begin{flushleft}
{\small
DARC: Present results from the DARC code  \\
ICFT: Results of  \cite{icft} \\ %Fern{\'a}ndez-Menchero et al. (2014) \\
}
\end{flushleft}
\end{table*}

For all ions the  lowest 98 levels  belonging to the 17  configurations (namely (1s$^2$) 2$\ell$2$\ell'$, 2$\ell$3$\ell'$ and 2$\ell$4$\ell'$) have been considered.  As stated earlier, for the calculations of $\Omega$ we adopted the DARC code. Specific details of the calculations are available in \cite{alx} for Al X, \cite{cl14} for Cl XIV, K XVI and Ge XXIX, and \cite{ti19} for Ti XIX. Briefly, all partial waves with angular momentum $J \le$ 40.5 have been considered, and to ensure convergence of $\Omega$ for all transitions and at all energies a ``top-up" based on  the Coulomb-Bethe  \citep{ab} and  geometric series  approximations was included for the allowed and forbidden transitions, respectively. Furthermore, values of $\Omega$ were determined up to a wide range of energies up to  380 Ryd (Al X),  660 Ryd (Cl XIV),  780 Ryd (K XVI), 1150 Ryd (Ti XIX) and 2500 Ryd (Ge XXIX). 

For the subsequent calculations of $\Upsilon$, resonances in a fine energy mesh (0.001 Ryd for most threshold regions) were resolved and averaged over  a  {\em Maxwellian} distribution of electron velocities. This distribution is commonly used and is  appropriate for most astrophysical applications. The density and importance of resonances for Be-like ions can be appreciated from figs. 6--11 of \cite{ti19} for a few transitions of Ti XIX. Results for $\Upsilon$ were obtained over a wide range of electron temperatures (T$_e$) fully covering that of their maximum fractional abundance in ionisation equilibrium  \citep*{pb}. Specifically,  $\Upsilon$ were reported up to log T$_e$ = 7.2 (Al X),  7.5 (Cl XIV),  7.5 (K XVI), 7.7 (Ti XIX) and 7.8 K (Ge XXIX). 

\cite{icft} adopted the AS code of \cite{as} to calculate energy levels and A-values. Their calculations are comparatively larger as they included 238 {\em fine-structure} levels of the (1s$^2$) 2(s,p) $n\ell$ ($n$ = 3--7,  $\ell$ = 0--4 for $n \le$5 and  $\ell$ = 0--2 for $n$ = 6--7) configurations. For most Be-like ions considered, their semi-relativistic approach generally yields comparable results with GRASP for energy levels and A-values. For example, in Table 1 we list the two sets of energies for the lowest 80 levels of Al X, where discrepancies are less than 0.02 Ryd and level orderings are also the same. Since detailed comparisons of energy levels with available  experimental and other theoretical results have already been made by us and \cite{icft}, we do not discuss these further. Similarly, we do not discuss the A-values, and rather focus on the more important parameter, i.e. $\Upsilon$.

\begin{figure*}
% \vspace{250pt}
\includegraphics[angle=90,width=0.9\textwidth]{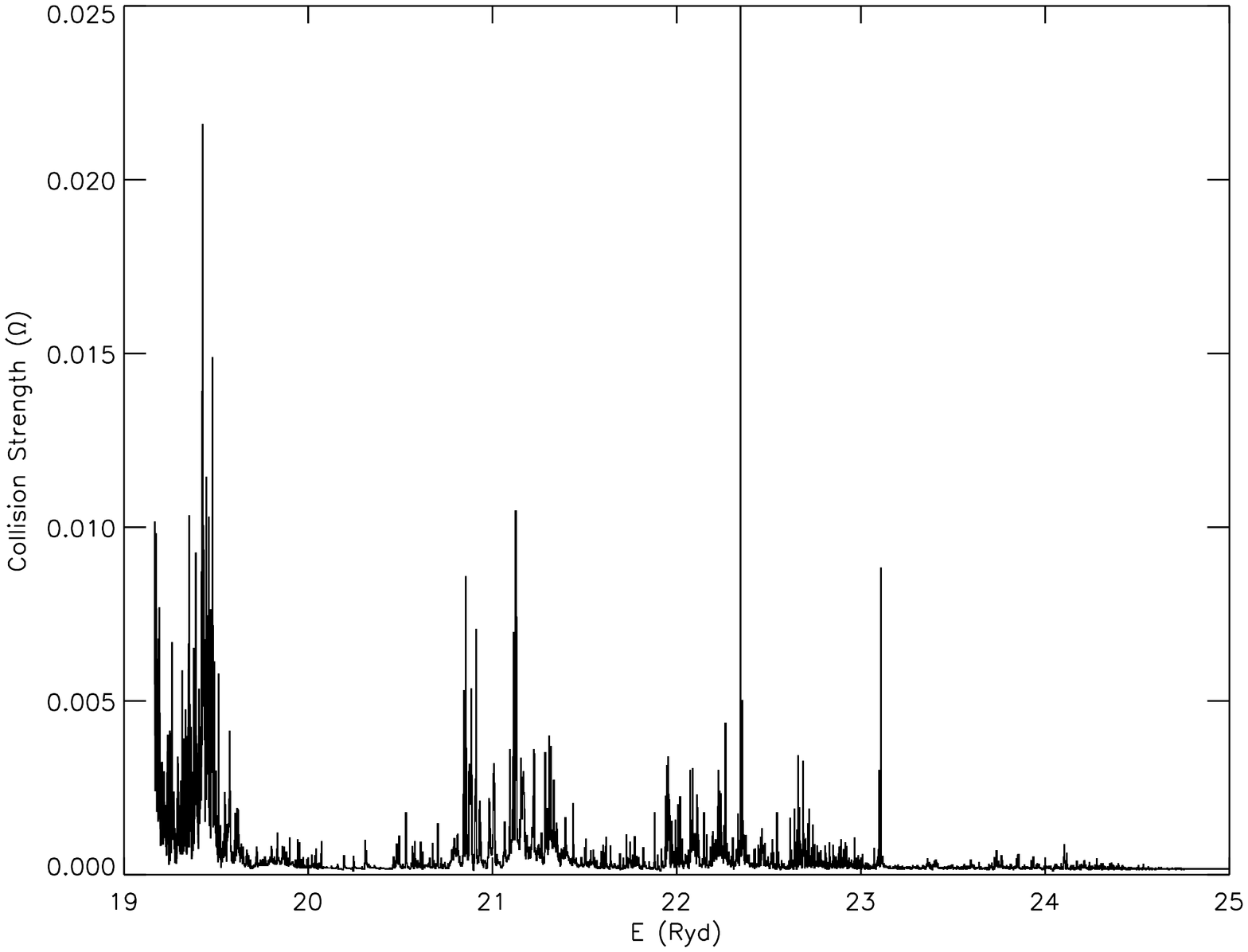}
%\includegraphics[width=\columnwidth]{fig1.eps}
% \vspace{9.5cm}
 \caption{Collision strengths for the 2s$^2$ $^1$S$_0$ -- 2p3p $^3$D$_{1}$ (1--26) transition of Al X.}
\end{figure*}

For the determination of $\Omega$, \cite{icft} obtained results for fine-structure transitions from the $LS$ calculations \citep{rm2} through the ICFT approach. They considered a slightly larger range of partial waves ($J \le$ 44.5) than ourselves. However, they only included electron {\em exchange}  up to $J$ =11.5 and for the rest performed a no-exchange calculation for expediency. This approach sometimes leads to sudden changes (at the overlap point) in the variation of $\Omega$ with $J$ -- see for example,  table 6 of \cite{o3}.  However, for most transitions it should not be a significant source of inaccuracy. More importantly, \cite{icft}  performed their calculations of $\Omega$ for  limited energy ranges, namely up to 90 Ryd (Al X), 165 Ryd (Cl XIV), 215 Ryd (K XVI), 300 Ryd (Ti XIX) and 680 Ryd (Ge XXIX),  lower by  a factor of $\sim$ 4  compared to our work. Unfortunately, such  energy ranges are  insufficient  \citep{ni11} to determine values of $\Upsilon$ at the high temperatures (up to $\sim$ 1.7$\times$10$^9$ K or equivalently $\sim$ 10,600 Ryd) for which these authors reported results. They did include high energy contributions to  $\Omega$ from the suggested formulae of \cite{bt}, but this approach, although computationally highly efficient, is perhaps a major source of inaccuracy, as discussed earlier by us for transitions of Fe XIV \citep{fe14}. Since we had already calculated values of $\Omega$ up to sufficiently high energies, there was no need for extrapolation to determine $\Upsilon$ for the ranges of T$_e$ reported by us. 

\cite{icft} resolved resonances in the threshold regions and averaged the values of $\Omega$ over a Maxwellian distribution of electron velocities to determine $\Upsilon$,   in a similar procedure to that employed by us. However, within thresholds they adopted a uniform mesh of 0.00001 z$^2$, where z is the {\em reduced} charge of the ion, i.e. Z--4. Consequently,  with increasing Z their adopted energy mesh becomes coarser. For example, for Al X  it is 0.00081 Ryd,  but for Ge XXIX is 0.0078 Ryd. Furthermore, this energy mesh was adopted only for partial waves with $J \le$ 11.5 (i.e. for the exchange calculations only), while for higher $J$  the mesh was coarser by a factor of 100, i.e. 0.001 z$^2$. By contrast, we adopted a uniform mesh ($\sim$ 0.001 Ryd) for {\em all} partial waves and for all ions.

\section{Effective collision strengths}

Since \cite{icft} have not reported results for $\Omega$ no direct comparisons with our work are possible. Therefore, we focus on a comparison of $\Upsilon$,  which are normally required for modelling applications. 

\subsection{Resonance transitions}

In Table 2 we list both sets of results for $\Upsilon$  for the resonance transitions of Al X among the lowest 80 levels which have the same orderings in both calculations, and at three electron temperatures, i.e. log T$_e$ = 4.3, 6.0 and 7.3 K. The first and the third are the lowest and the highest {\em  common} temperatures between the two calculations, whereas the second is the most relevant for modelling applications, because 10$^{6.1}$ K is the temperature at which Al X has its maximum abundance in ionisation equilibrium  \citep{pb}.

\begin{figure*}
% \vspace{250pt}
\includegraphics[angle=-90,width=0.9\textwidth]{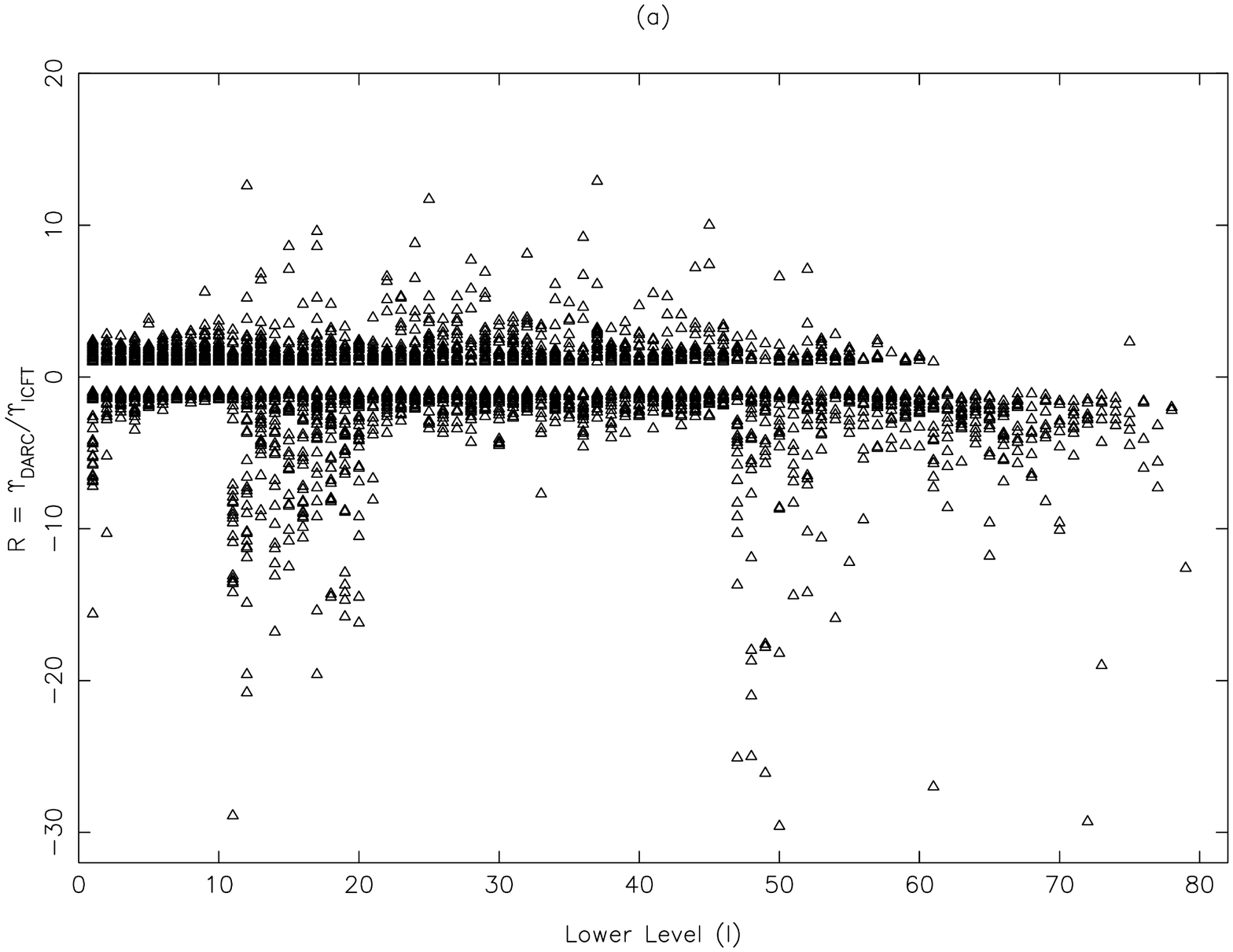}
\includegraphics[angle=-90,width=0.9\textwidth]{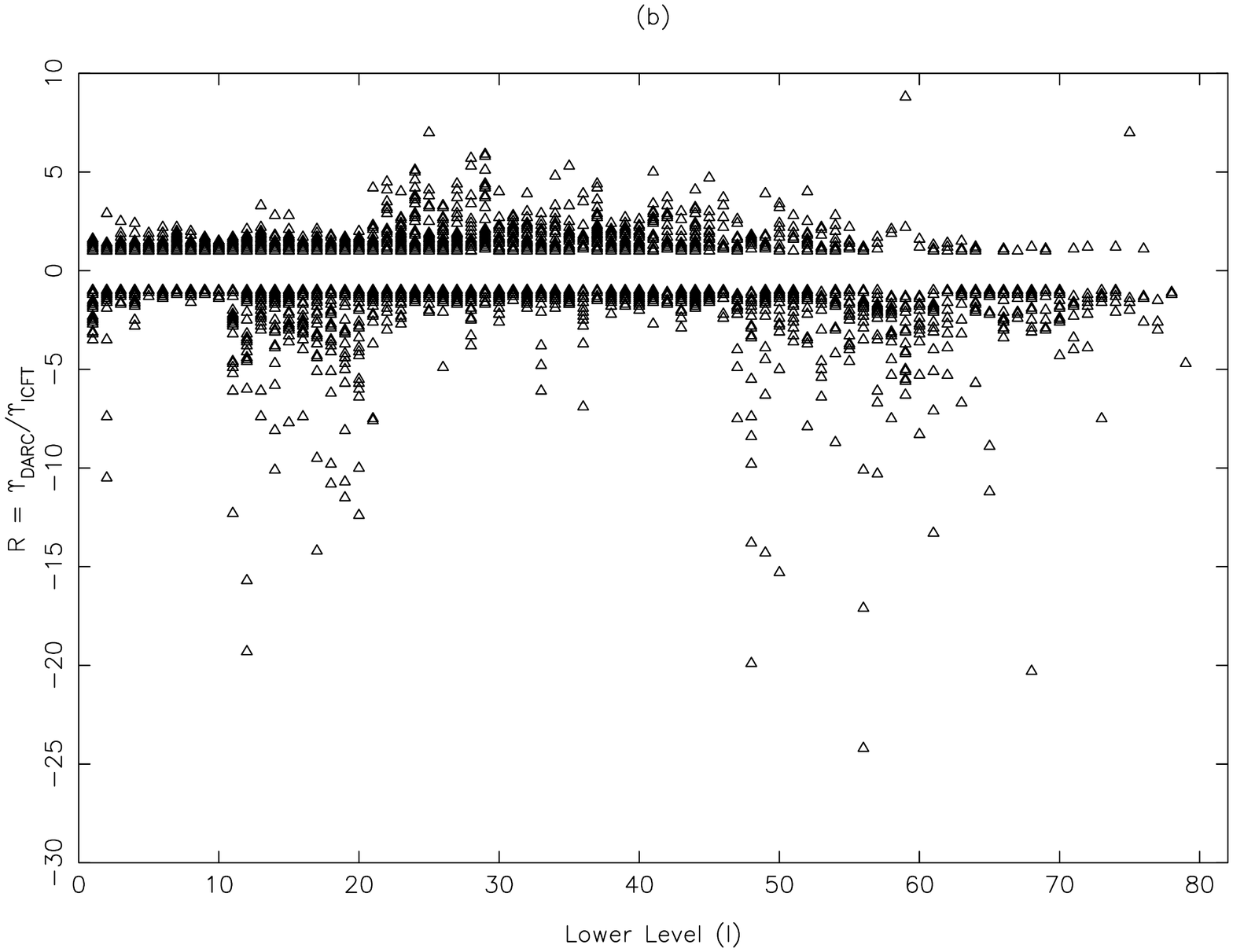}
\end{figure*}

%\newpage  
\begin{figure*}
% \vspace{250pt}
\includegraphics[angle=-90,width=0.9\textwidth]{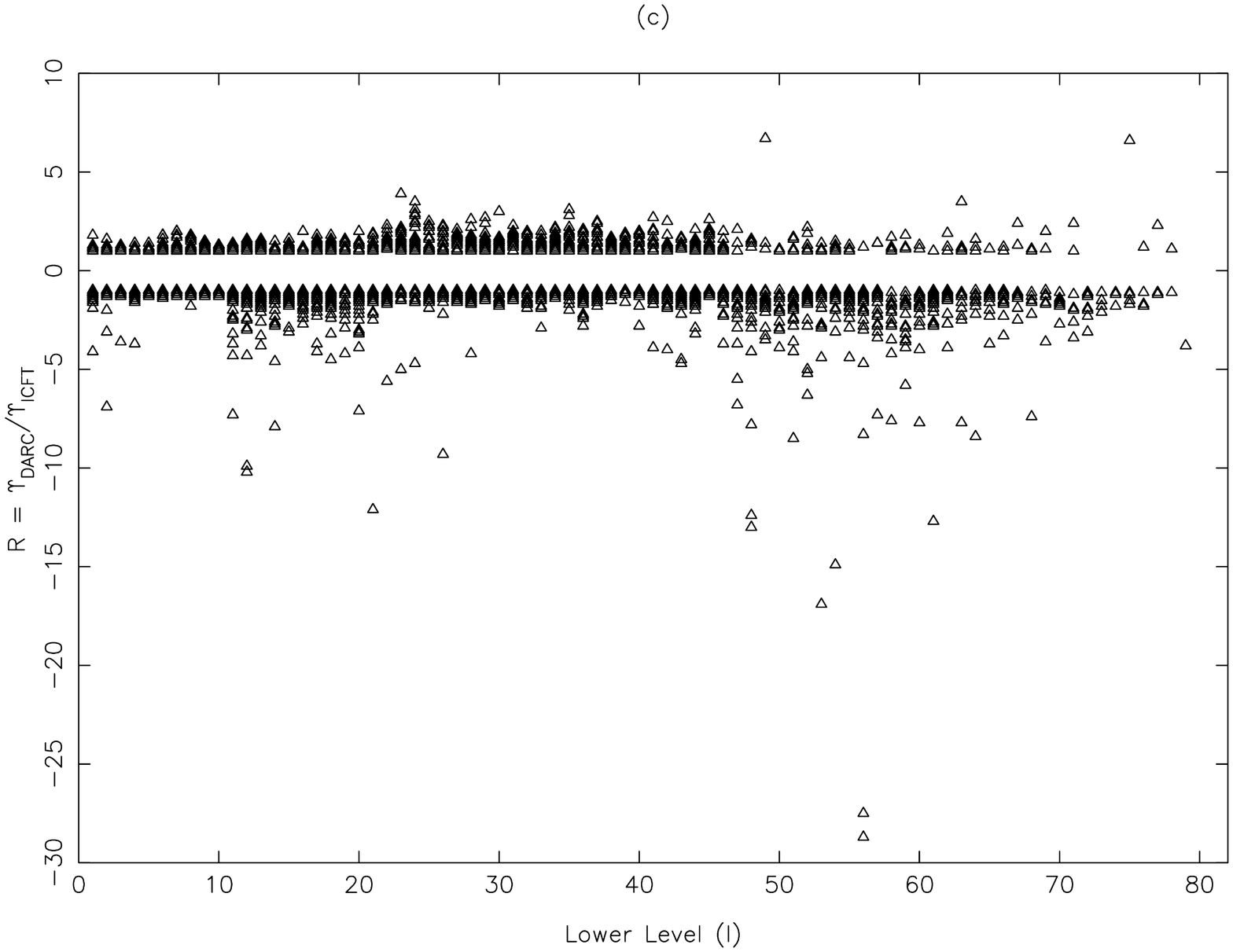}
%\includegraphics[width=\columnwidth]{fig1.eps}
% \vspace{9.5cm}
 \caption{Ratio (R) of  DARC and ICFT effective collision strengths for transitions among the lowest 80 levels of  Al X, at electron temperatures of (a) T$_e$ = 10$^{4.3}$ K, (b)  T$_e$ = 10$^{6.0}$ K and  (c) T$_e$ = 10$^{7.3}$ K. Negative R values indicate that $\Upsilon_{\rm ICFT}$ $>$ $\Upsilon_{\rm DARC}$. }
\end{figure*}

For most transitions listed in Table 2, the $\Upsilon$ of \cite{icft} are larger, by up to a factor of 15, at all temperatures, particularly the lower ones. However, for some transitions our results are higher by up to a factor of 2, such as 1--26/27/28 (i.e. 2s$^2$ $^1$S$_0$ -- 2p3p $^3$D$_{1,2,3}$). These transitions are {\em forbidden} in both the LS and $jj$ coupling schemes, and hence resonances for these are significant (particularly towards the lower end of the energy range)  as shown in Fig. 1 for  2s$^2$ $^1$S$_0$ -- 2p3p $^3$D$_{1}$ (1--26). We note that even a  slight shift in resonance positions can affect the calculated values of $\Upsilon$, particularly at low(er) temperatures. However, as shown in Table 1 the two sets of energies obtained by the AS and GRASP codes are comparable for most levels, including  2p3p $^3$D$_{1,2,3}$. If these near threshold resonances are missing from the calculations of \cite{icft} then their results for $\Upsilon$ will clearly be lower.

Among the transitions for which $\Upsilon_{\rm ICFT}$  are larger than $\Upsilon_{\rm DARC}$  at  log  T$_e$ = 4.3 K, are 1--64 (2s$^2$ $^1$S$_0$ -- 2p4s $^1$P$^o_1$) and 1--80 (2s$^2$ $^1$S$_0$ -- 2p4f $^1$F$_3$), which are allowed and forbidden, respectively. The 1--64 transition is  weak with f = 5$\times$10$^{-4}$ in our work (GRASP) and f = 1$\times$10$^{-4}$ from AS, and yet $\Upsilon_{\rm ICFT}$  $>$  $\Upsilon_{\rm DARC}$ by a factor of $\sim$ 6  at the lowest common temperature, while $\Upsilon_{\rm ICFT}$ is smaller by a factor of 2 at log  T$_e$ = 7.3 K. For this transition,  resonances are not  prominent as expected. Similarly, for the 1--80 (forbidden) transition we do not observe  any significant resonances, although \cite{icft}  may do so, because their calculations include 238 levels in comparison to  only 98 in ours. Therefore, we now focus our attention on the highest temperature of 10$^{7.3}$ K (equivalent to $\sim$ 126 Ryd) at which the contributions of resonances, if any, will not be appreciable.

Among the transitions listed in Table 2, the one which shows the largest factor (4) by  which  $\Upsilon_{\rm ICFT}$ is higher than $\Upsilon_{\rm DARC}$ is 1--70 (2s$^2$ $^1$S$_0$ -- 2p4p $^3$P$_0$). In fact, the difference between the two calculations increases with increasing temperature (from a factor of 2.6 at T$_e$ = 10$^{4.3}$ K to 4.1 at  10$^{7.3}$ K). However, this is a forbidden transition for which $\Omega$ decreases (or becomes nearly constant) with increasing energy -- see table 4 of \cite{alx}. Therefore, the behaviour of $\Upsilon$ in the calculations of \cite{icft} is difficult to understand for some transitions.

\subsection{All transitions}

The comparisons of $\Upsilon$ shown in Table 2 are for a very limited range of transitions. Therefore, in Fig. 2 (a, b and c) we compare the two sets of $\Upsilon$ for all 3160 transitions  among the 80 levels of Al X at  three temperatures of Table 2. It is clear from these figures that for a majority of transitions the $\Upsilon$ of \cite{icft} are significantly larger (by up to a factor of 30) at all temperatures. More specifically, for  69\% and 42\% of the transitions the two sets of $\Upsilon$ differ  by over 20\% at the lowest and the highest common temperatures of 10$^{4.3}$ and 10$^{7.3}$ K, respectively. Also,  for a majority of transitions the values of $\Upsilon_{\rm ICFT}$ are larger. A similar comparison is found for transitions in other ions, namely Cl XIV, K XVI, Ti XIX and Ge XXIX. 

\subsection{$n$ = 2 transitions}

We now discuss the discrepancies between the two sets of $\Upsilon$ for specific range of transitions. Among the lowest 10 levels are 45 transitions belonging to the $n$ = 2 configurations. At T$_e$ = 10$^{4.3}$ K, for $\sim$ 30\% of transitions discrepancies are up to a factor of 2.5, largest for the 4--6: 2s2p $^3$P$^o_2$ -- 2p$^2$ $^3$P$_0$, which is forbidden. For half the transitions $\Upsilon_{\rm DARC}$ $>$ $\Upsilon_{\rm ICFT}$  and for the other half $\Upsilon_{\rm ICFT}$  $>$  $\Upsilon_{\rm DARC}$, and there are no systematics because both types  (allowed and forbidden)   show the discrepancies. This may be due to the position of resonances, as discussed in section 3.1, because at T$_e$ = 10$^{6.0}$  and 10$^{7.3}$ K both sets of $\Upsilon$ agree within $\sim$20\%.

\subsection{$n$ = 3 transitions}

There are 630 transitions among the $n$ = 3 configurations belonging to the levels between 11 and 46. At T$_e$ = 10$^{4.3}$  and 10$^{6.0}$ K for about half the transitions (including both allowed and forbidden)  $\Upsilon_{\rm DARC}$ $>$ $\Upsilon_{\rm ICFT}$, whereas a third of transitions have $\Upsilon_{\rm ICFT}$  $>$  $\Upsilon_{\rm DARC}$ at T$_e$ = 10$^{7.3}$ K. It is difficult to explain these discrepancies at all temperatures, particularly for the allowed transitions.
\subsection{$n$ = 4 transitions}

These transitions belong to levels 47 and higher and are 561 in number. For a majority of transitions (about two third) $\Upsilon_{\rm ICFT}$  $>$  $\Upsilon_{\rm DARC}$ at {\em all} temperatures.  Again, these large discrepancies are difficult to understand, particularly at T$_e$ = 10$^{7.3}$ K and for allowed transitions. Nevertheless, we will return to the source of  discrepancies in section 4, apart from those already discussed in section 2.

\subsection{Strong allowed transitions}

We now discuss the {\em strong} allowed transitions, i.e. with f $\ge$ 0.1. Among the lowest 80 levels of Al X given in Table 1, there are 79 such transitions. At the most relevant temperature of Al X (i.e. 10$^6$ K) both sets of $\Upsilon$ agree within $\sim$20\% for all transitions, but for about a third $\Upsilon_{\rm ICFT}$  $>$  $\Upsilon_{\rm DARC}$ at T$_e$ = 10$^{4.3}$  and 10$^{7.3}$ K. Discrepancies are comparatively larger at the lowest common temperature and many transitions belong to the lowest 46 levels of the $n$ = 2 and 3 configurations.

\subsection{1--3 transition of P XII}

Finally, we note that the comparison shown by \cite{icft} in their fig. 5 with the $\Upsilon$ of \cite{fpk} for the 2s$^2$ $^1$S$_0$ -- 2s2p $^3$P$^o_1$ (1--3) transition of P XII is incorrect. For this transition there is no appreciable difference between the $\Upsilon$ interpolated by \cite{fpk} and those calculated by \cite{icft}. In fact,  for this transition the interpolated values for other ions are underestimated (by up to a factor of two), rather than overestimated as shown by \cite{icft}.  This can be seen from table 6 of \cite{alx} for Al X, and tables 16 and 17 of \cite{cl14} for Cl XIV and K XVI. The error in calculating $\Upsilon$ has occurred by \cite{icft} because they have mistakenly taken the coefficient $a$1 to be positive rather than negative ($a$1 = --0.026314), as  given by \cite{fpk} in his table II.

\section{Conclusions}
 In this paper we have compared two sets of electron impact excitation effective collision strength ($\Upsilon$) for transitions in Be-like ions obtained by the $R$-matrix method in  semi and fully relativistic approaches, i.e. ICFT  and DARC. Both approaches should provide comparable results for a majority of transitions. However, significant differences of up to more than an order of magnitude are noted for at least 50\% of the transitions of all ions with 13 $\le Z \le$ 32, and over the entire range of electron temperature. In most cases the $\Upsilon$ from ICFT are significantly larger than those obtained with DARC. We believe the discrepancies have arisen mainly due to some compromises made by \cite{icft} in calculating $\Omega$ and subsequently $\Upsilon$, as noted in section 2.  Similar large discrepancies between the two independent $R$-matrix approaches have also been noted in the past -- see for example, \cite{fe14} for Fe XIV, \cite{he5} and references therein for He-like  and \cite{li1} for Li-like ions. Therefore, it appears that the implementation of the ICFT approach, although computationally highly efficient (hence allowing data for many ions to be produced over relatively short periods),  may not be completely robust. Indeed,  this has also been confirmed by \cite*{ssb} in their calculations for O III, who noted that for some transitions the ICFT results can be significantly overestimated in comparison to the Breit-Pauli (or other similar approaches, such as DARC). Therefore, we recommend that the excitation rates reported by \cite{icft} should be used with caution and a re-examination of their results would be helpful.

As already stated in section 1, assessing the accuracy of $\Upsilon$ is a difficult task \citep{fst}, mainly because large calculations cannot be easily reproduced. However, when two (or more) calculations of comparable complexity and with similar approaches (such as $R$-matrix) become available, the discrepancies observed are often striking, as noted above for He-like, Li-like and Be-like  ions. Therefore, the true accuracy of any atomic data can only be assessed either by modelling  applications or the comparisons, as shown in this work. However, since observational data are generally limited, a comparison with theoretical results is often inconclusive. The large discrepancies in $\Upsilon$ observed here for a majority of transitions of many Be-like ions do create suspicion in the minds of users. Nevertheless, as discussed in our published papers, our accuracy assessment for a majority of transitions in the temperature ranges of interest remains the same, i.e. $\sim$20\%. Although no accuracy assessment has been made by \cite{icft}, their reported data would appear to be less reliable due to  the compromises made in their calculations particularly in the chosen energy mesh and the energy ranges over which values of $\Omega$ were calculated before extrapolation.  However,  we also stress that there is  scope  for improvement in our work (as there is for any calculation), especially  for transitions involving  levels of the $n$ = 4 configurations. The inclusion of  levels of the $n \ge$ 5 configurations  in the collisional calculations may improve the reported values of $\Upsilon$ due to the additional resonances arising.  However, until then we believe that our reported results  for radiative and excitation rates for transitions in  Be-like ions are (probably) the most exhaustive and accurate available to date, and should be useful for the modelling of astrophysical plasmas.

\section*{Acknowledgments}

KMA is thankful to  AWE Aldermaston for    financial support  and we thank an anonymous referee for suggesting discussions about sections 3.3--3.6.


\begin{thebibliography}{99}
\bibitem[\protect\citeauthoryear{Aggarwal \& Hibbert}{1991}]{o3}
Aggarwal K. M.,   Hibbert A., 1991, J. Phys. B, 24, 3445
\bibitem[\protect\citeauthoryear{Aggarwal \& Keenan}{2008}]{ni11}
Aggarwal K. M.,  Keenan F. P., 2008,  Eur. Phys. J.,  D 46,  205 
\bibitem[\protect\citeauthoryear{Aggarwal \& Keenan}{2012a}]{li1}
Aggarwal K. M.,   Keenan F. P.,  2012a, At. Data Nucl. Data Tables, 98, 1003 
\bibitem[\protect\citeauthoryear{Aggarwal \& Keenan}{2012b}]{ti19}
Aggarwal K. M.,   Keenan F. P.,  2012b, Phys. Scr., 86, 055301 
\bibitem[\protect\citeauthoryear{Aggarwal \& Keenan}{2013a}]{fst}
Aggarwal K. M.,   Keenan F. P.,  2013a, Fusion Sci. Tech, 63, 363 
\bibitem[\protect\citeauthoryear{Aggarwal \& Keenan}{2013b}]{he5}
Aggarwal K. M.,   Keenan F. P.,  2013b, Phys. Scr., 87, 045304
\bibitem[\protect\citeauthoryear{Aggarwal \& Keenan}{2014a}]{alx}
Aggarwal K. M.,   Keenan F. P.,  2014a, MNRAS, 438, 1223 
\bibitem[\protect\citeauthoryear{Aggarwal \& Keenan}{2014b}]{fe14}
Aggarwal K. M.,   Keenan F. P.,  2014b, MNRAS, 445, 2015 
\bibitem[\protect\citeauthoryear{Aggarwal \& Keenan}{2014c}]{cl14}
Aggarwal K. M.,   Keenan F. P.,  2014c, Phys. Scr., 89, 125401
\bibitem[\protect\citeauthoryear{Badnell}{1997}]{as}
Badnell N. R.,  1997, J. Phys. B, 30, 1
\bibitem[\protect\citeauthoryear{Berrington, Eissner \& Norrington}{Berrington et al.}{1995}]{rm2}
Berrington K. A., Eissner W. B., Norrington P. H., 1995,        Comput. Phys. Commun.,  92,  290
\bibitem[\protect\citeauthoryear{Bryans, Landi \& Savin}{Bryans et al.}{2009}]{pb}
Bryans P., Landi E.,    Savin D. W., 2009,  ApJ,  691, 1540 
\bibitem[\protect\citeauthoryear{Burgess \& Sheorey}{1974}]{ab}
Burgess A.,   Sheorey V. B.,  1974, J. Phys.,  B7,  2403 
\bibitem[\protect\citeauthoryear{Burgess \& Tully}{1992}]{bt}
Burgess A.,    Tully J. A., 1992, A\&A,   254, 436 
%\bibitem[\protect\citeauthoryear{Feldman \& Seely}{1985}]{fs}
%Feldman U., Seely J. F., 1985, At. Data Nucl. Data Tables, 32, 305 
\bibitem[\protect\citeauthoryear{Fern{\'a}ndez-Menchero, Del Zanna \& Badnell}{Fern{\'a}ndez-Menchero et al.}{2014}]{icft}
 Fern{\'a}ndez-Menchero L., Del Zanna G.,  Badnell N.R., 2014, A\&A, 566, A104
\bibitem[\protect\citeauthoryear{Grant et al.}{1980}]{grasp0}
Grant I. P., McKenzie B. J.,  Norrington P. H.,  Mayers D. F.,   Pyper N. C., 1980,	Comput. Phys. Commun.,  21,  207  
\bibitem[\protect\citeauthoryear{Keenan}{1988}]{fpk}
Keenan F. P., 1988, Phys. Scr., 37, 57 
\bibitem[\protect\citeauthoryear{Landi et al.}{2001}]{el1}
Landi E., Doron R., Feldman U., Doscheck G. A., 2001, ApJ, 556, 912 
\bibitem[\protect\citeauthoryear{Liang et al.}{2010}]{gyl}
Liang G. Y., Badnell N. R., Crespo L{\' o}pez-Urrutia J. R., Baumann T. M., Del Zanna G., Storey P. J., Tawara H., Ullrich J., 2010, ApJS, 190, 322
\bibitem[\protect\citeauthoryear{Storey, Sochi \& Badnell}{Storey et al.}{2014}]{ssb}
Storey P. J., Sochi T., Badnell N. R.,  2014, MNRAS, 441, 3028

\end{thebibliography}
\end{document}